# High-Q silica zipper cavity for optical radiation pressure driven MOMS switch

Tomohiro Tetsumoto and Takasumi Tanabe[a]

Department of Electronics and Electrical Engineering, Faculty of Science and Technology,

Keio University, 3-14-1 Hiyoshi Kohoku, Yokohama 223-8522, Japan

We design a silica zipper cavity that has high optical and mechanical $Q$ (quality factor) values and demonstrate numerically the feasibility of a radiation pressure driven micro opto-mechanical system (MOMS) directional switch. The silica zipper cavity has an optical $Q$ of $6.0 \times 10^4$ and an effective mode volume $V_{mode}$ of $0.66\lambda^3$ when the gap between two cavities is 34 nm. We found that this $Q/V_{mode}$ value is five times higher than can be obtained with a single nanocavity design. The mechanical $Q$ ($Q_m$) is determined by thermo-elastic damping and is $2.0 \times 10^6$ in a vacuum at room temperature. The opto-mechanical coupling rate $g_{OM}$ is as high as 100 GHz/nm, which allows us to move the directional cavity-waveguide system and switch 1550-nm light with 770-nm light by controlling the radiation pressure.

## I. INTRODUCTION

Cavity opto-mechanics enables the dynamic manipulation of the oscillation of cavity structures[1]. Opto-mechanics is now attracting considerable attention, and there are a number of studies that discuss aspects of its fundamental physics such as ground state cooling[2,3], electromagnetically induced transparency[4,5] and quantum information processing[6,7]. On the other hand, studies related to practical applications have just begun. Preliminary experiments have been reported on resonant wavelength conversion[8,9], mass-sensing[10], and an accelerometer[11], but the number of applications is still very limited. Photonic crystal nanocavities[12,13], which are used in various types of nonlinear optical research[14, 15], are attractive candidates as platforms for cavity opto-mechanics applications owing to their strong light confinement and extremely small mode volume. A number of studies have been performed using different types of photonic crystals such as in-plane two-dimensional photonic crystals[16, 17], doubled layer photonic crystal slabs[18], coupled nanobeam cavities[19], and single nanobeam cavitiy[5]. The key to fabricating a good opto-mechanical device is to localize the optical and mechanical fields in the same place and make their quality factors ($Q$s) high.

---
[a] takasumi@elec.keio.ac.jp

The mechanical $Q$ is limited by, for example, viscous or fluidic damping, thermo-elastic damping, and support-induced damping[20]. In particular, viscous damping is so high that most experiments are usually carried out in a vacuum where thermo-elastic loss is usually the dominating factor[21]. Silica has a very low thermal conductivity (long thermal relaxation time) and a low thermal expansion coefficient, so it is potentially a material with a low thermal loss. However, there have been few studies on fabricating silica photonic crystals because silica has a low refractive index. In this paper, we design a silica zipper cavity with high optical and mechanical $Q$ values even at room temperature and analyze its optical and mechanical properties in detail.

This paper is organized as follows; first in Sec. II we describe the design and the optical properties of our silica zipper cavity. Usually, a high $Q$ is difficult to achieve with a low-index material, but we show that our zipper cavity even exhibits a high $Q/V$. Then in Sec. III, we report the mechanical characteristics of our silica zipper cavity. Section IV discusses the numerical results we obtained for radiation pressure driven optical switching. We finish with a conclusion.

## II. OPTICAL PROPERTIES OF SILICA ZIPPER CAVITY

### A. High-*Q* design

A zipper cavity is composed of two nanobeam cavities and the light is localized and enhanced in the gap between them[19, 22]. This localized electromagnetic field is strongly dependent on the mechanical displacement of the cavity; hence strong opto-mechanical coupling can be obtained. To increase the opto-mechanical coupling, the nanobeam cavity must have a high optical *Q*. Although there have been studies of high *Q* nanobeam cavities with high refractive index materials[23, 24, 25], it is a challenge to fabricate a high-*Q* cavity with low reflective index materials. Some studies have reported the fabrication of high-*Q* nanobeam cavities in low refractive index materials such as silica[26] ($n = 1.46$), silicon nitride[27] ($n = 2.0$) and polymer[28] ($n = 1.34$). So, we follow a previously reported nanobeam cavity design[26], but apply it to a zipper design[22].

A schematic diagram of a zipper cavity is shown in FIG. 1(a). Two nanobeam cavities with width *w*, slab thickness *t*, hole height $h_x$, and hole width $h_y$, are positioned with slot gap *s*. We define the lattice constant between the *m*'th and (*m*+1)'th holes as $a_m$.

We first design a nanobeam cavity. To form a potential and confine the light, we design the lattice constant around the center of the cavity to be slightly smaller than that of the edge holes, which have a lattice constant of *a*. Lattice constant versus hole number is plotted in FIG. 1(b). We calculate the one-dimensional band diagram of a single silica nanobeam cavity with the parameters $w = 2.6a$, $t = 1.1a$, $h_x = 0.5a$, and $h_y = 0.7w$ using the MIT photonic band software package[29] (FIG. 1(c)). The band edge of the 1D photonic crystal with a $0.92a$ lattice constant (red curve) is at a higher frequency and is within the photonic bandgap of a $1.0a$ lattice (blue curve). Hence, the light can be localized by modulating the lattice constant locally to $0.92a$ for the cavity region. We calculate the mode profile and the *Q* by using the three-dimensional finite-difference time-domain (FDTD) method (Meep FDTD package)[30]. We set the total number of holes at 53 including 15 defect holes and set the lattice constant at $a = 335$ nm, which enables us to design the resonant wavelength of the fundamental bonded-mode at around 770 nm. The $Q$ for a single silica nanobeam cavity with the parameters described above is $Q = 1.0 \times 10^4$, and $V_{mode} = 0.60\, \lambda^3$.

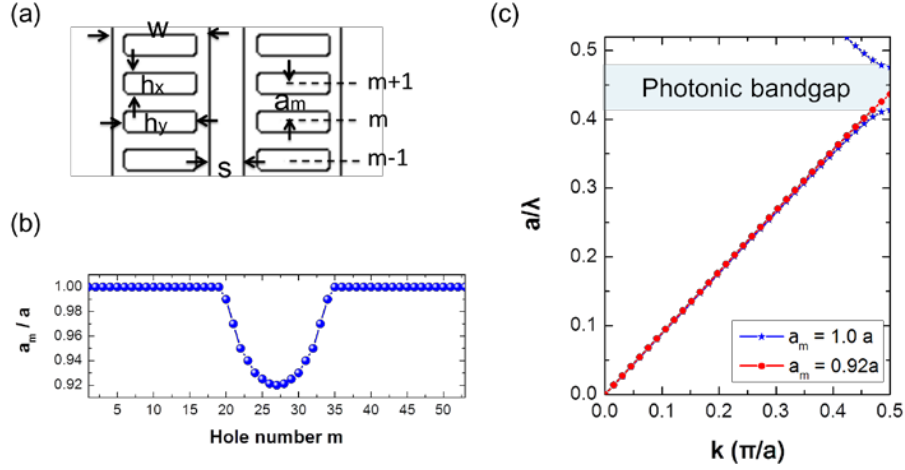

FIG. 1: (a) Schematic illustration of a zipper cavity. (b) Lattice constant as a function of hole number along the cavity. (c) 1D band diagram of the TE polarized mode of a single silica nanobeam cavity.

We next designed the zipper cavity. We again set the lattice constant at $a = 335$ nm. We made the cavity smaller and more sensitive to the slot gap by designing the resonant wavelength in the visible region. The zipper cavity has two supermodes that have strong opto-mechanical coupling. In accordance with previous work[22], we refer to an even parity supermode as a bonded mode, and an odd parity supermode as an anti-bonded mode. We compute the electric mode profile of the fundamental bonded and anti-bonded modes by using 3D FDTD (FIG. 2(a)). A typical field intensity map along the $y$ direction is shown in FIG. 2(b). The bonded mode has its maximum intensity in the air slot gap. On the other hand, the anti-bonded mode has its maximum intensity in the dielectric cavity. These features cause the difference in behavior of these two modes.

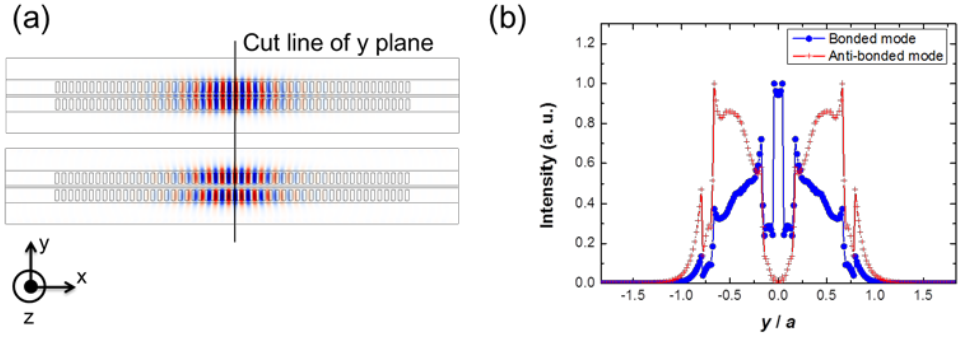

FIG. 2: (a) Electric field amplitude distribution of zipper cavity. The bonded mode is at the top and the anti-bonded mode below it. (b) Electric intensity distribution across the center of the zipper cavity of gap $s = 0.3a$. The intensity is normalized with the maximum intensity of the field and the y-axis is normalized with lattice constant $a$.

Then, we calculate the cavity $Q$ and the effective mode volume as a function of the gap $s$. The results are shown in FIG. 3. The mode volume of an anti-bonded mode does not depend strongly on the gap, while it has a higher $Q/V_{mode}$ than the bonded mode in most ranges. This is because the anti-bonded mode field is confined in the dielectric region. The highest $Q/V_{mode}$ is obtained when the gap $s = 268$ nm, $Q = 6.7 \times 10^4$, and $V_{mode} = 1.12\lambda^3$. In contrast, the bonded mode exhibits a high $Q/V_{mode}$ as the gap becomes smaller because of the enhanced light localization in the slot gap. However, when the gap approaches zero, the light confinement is slightly weakened and $Q/V_{mode}$ decreases. The maximum $Q/V_{mode}$ is obtained when the gap $s = 34$ nm, $Q$ is $6.0 \times 10^4$ and $V_{mode}$ is $0.66\lambda^3$. We would like to emphasize that this $Q/V_{mode}$ value is about five times larger than that obtained with a single silica nanobeam cavity. It is interesting to note that a higher $Q/V_{mode}$ can be obtained by using coupled nanobeam cavities instead of one. This finding is useful not only for opto-mechanics applications but also for cavity quantum electrodynamics applications, which often require a high Purcell factor[31].

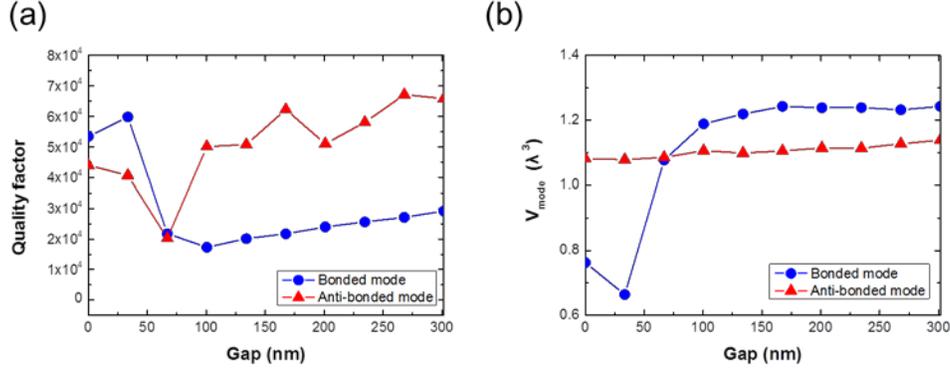

FIG. 3: (a) Gap dependence versus quality factor. (b) Gap dependence versus effective mode volume.

## B. Opto-mechanical coupling

We use the 3D FDTD to obtain the resonant wavelength for different slot gaps. The result is shown in FIG. 4(a). The bonded mode exhibits a strong dependence on the gap change, but the dependence of the anti-bonded mode is weak. The difference in the behavior of these two modes can be understood from the difference between the mode profiles as shown in FIG. 2. Since the gap change has an adiabatic dependence, we can calculate both the opto-mechanical coupling ratio $g_{OM}/2\pi = d\omega/ds$ and the induced optical radiation force $F = -dU/ds = -\hbar d\omega/ds$ (FIG. 3(b)). We obtain a strong opto-mechanical coupling ratio for a bonded mode of larger than 100 GHz/nm when the gap $s = 34$ nm. This value is sufficient to move the structure by radiation force as we discuss in Sec. IV.

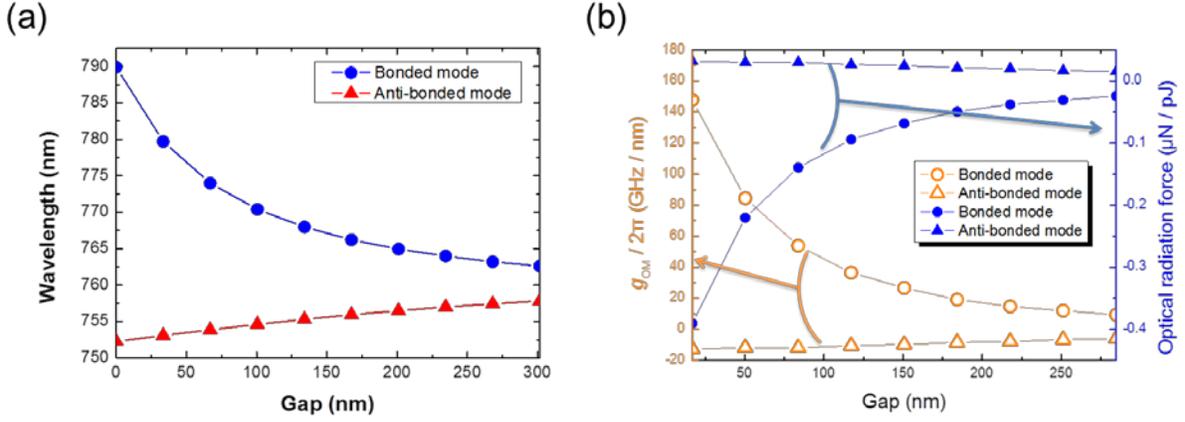

FIG. 4: (a) Gap dependence versus cavity resonant wavelength. (b) Gap dependence versus opto-mechanical coupling rate and optical radiation force. Orange lines show the opto-mechanical coupling rate and blue lines show the induced optical radiation force. A positive sign indicates repulsive force and a negative sign indicates attractive force.

### III. MECHANICAL PROPERTIES OF SILICA ZIPPER CAVITY

Identifying the oscillation mode of the cavity is important if we are to understand the opto-mechanical properties of the device. The zipper cavity has four main oscillation modes; the in-plane horizontal mode, out-of-plane vertical mode, twisting mode, and compression mode. They can be categorized into two oscillation types due to the difference between the oscillation phases of the two cavities; the common mode and differential mode. We named the mechanical modes using a label that consists of three parts; the first character indicates the oscillation type (h: horizontal mode, v: vertical mode, t: twisting mode, c: compression mode), the second character is the mode dimension, and the third character is the oscillation phase (c: common mode, d: differential mode). As regards opto-mechanics in a zipper cavity, we focus on the $h_{1d}$ mode, since it is the dominant mode that can be excited by radiation pressure.

We calculate the mechanical properties of the zipper cavity using the finite element method (FEM) with COMSOL Multiphysics[32]. We assume room temperature and vacuum conditions. We prepare a zipper structure with a total length of 18.3 μm, a total hole number of 53, where $a$ = 335 nm,

$w = 871$ nm, $t = 369$ nm, $h_x = 168$ nm, $h_y = 610$ nm, and $s = 101$ nm. The calculated mechanical frequency of the fundamental horizontal differential mode is $\Omega_m/2\pi = 13.5$ MHz with silica material properties; density $\rho = 2203$ kg/m$^3$, Young's modulus $E = 73.1$ GPa, and Poisson's ratio $\mu = 0.17$. We calculate the effective mass and effective spring constant using the following equations.

$$m_{eff} = \frac{\rho \iiint u^2 dV}{u_{max}^2} \quad (1)$$

$$k_{eff} = \Omega_m^2 m_{eff} \quad (2)$$

where $u$ indicates the displacement of the structure. The results are summarized in Table I. We obtained an effective mass of $m_{eff} = 5.37$ pg for the h$_{1d}$ mode.

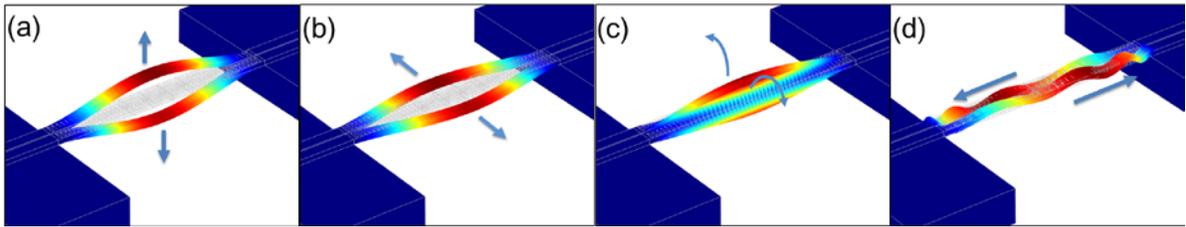

FIG. 5: Differential mechanical oscillation of (a) vertical, (b) horizontal, (c) twisting (d) compression modes.

TABLE I. Summary of mechanical mode properties of zipper cavity.

| Mode label | $\Omega_m/2\pi$ (MHz) | $m_{eff}$ (pg) | $k_{eff}$ (N/m) |
|---|---|---|---|
| $v_{1c}$ | 5.30 | 4.93 | 5.46 |
| $v_{1d}$ | 5.30 | 4.93 | 5.47 |
| $v_{2c}$ | 14.4 | 5.30 | 43.6 |
| $v_{2d}$ | 14.4 | 5.30 | 43.7 |
| $v_{3c}$ | 28.2 | 5.46 | 172 |
| $v_{3d}$ | 28.2 | 5.46 | 172 |
| $h_{1d}$ | 13.5 | 5.37 | 38.8 |
| $h_{1c}$ | 13.6 | 5.36 | 38.9 |
| $h_{2d}$ | 33.8 | 5.84 | 264 |
| $h_{2c}$ | 33.9 | 5.84 | 265 |
| $h_{3d}$ | 61.2 | 5.51 | 814 |
| $h_{3c}$ | 61.2 | 5.50 | 814 |
| $t_{1d}$ | 43.5 | 2.26 | 168 |
| $t_{1c}$ | 43.5 | 2.26 | 168 |
| $c_{1c}$ | 116 | 6.50 | 3460 |
| $c_{1d}$ | 117 | 6.47 | 3500 |

It is known that thermo-elastic loss is usually the dominant mechanical loss in a vacuum and at room temperature[21, 33, 34]. The loss is expressed by the following equation.

$$Q_m = \frac{c_p \rho}{E\alpha^2 T} \frac{1+\Omega_m^2 \tau_{th}^2}{\Omega_m \tau_{th}} \quad (3)$$

where $c_p$ is the specific heat, $\alpha$ is the thermal expansion coefficient, $T$ is the atmospheric temperature, $\tau_{th}$ is the thermal relaxation time defined by $\tau_{th} = c_p \rho w^2 (\pi^2 \kappa_{th})^{-1}$ where $\kappa_{th}$ is the thermal conductivity, and $w$ is the width of the beam in the plane of vibration. We calculate $\tau_{th}$ and $Q_m$ with the following parameters; $c_p = 780$ J·kg$^{-1}$·K$^{-1}$, $\alpha = 0.57 \times 10^{-6}$ K$^{-1}$, $\kappa_{th} = 1.4$ W·m$^{-1}$·K$^{-1}$ and room temperature $T = 297$ K. The obtained $\tau_{th} = 94$ ns and $Q_m = 2.0 \times 10^6$. This large mechanical quality factor is due to the low thermal conductivity and low thermal expansion coefficient of silica. Although a large $Q_m$ is attractive for some opto-mechanics applications such as the regenerative amplification of mechanical modes, for our particular switching application, we do not require a high $Q_m$, since we do not want the structure to oscillate for a long time. It should be noted that an important parameter for an efficient opto-mechanical switching operation is a large $g_{OM}$, where $Q_m$ is independent of this value. Although we performed our previous calculation in a vacuum, air will yield additional damping that will reduce $Q_m$. As a result we will reach a steady state more quickly.

## IV. RADIATION PRESSURE ASSISTED OPTO-MECHANICAL SWITCH

In this section, we propose an opto-mechanical directional coupler switch based on a silica zipper cavity. A directional coupler is a passive device that can route the direction of light. It is composed of two evanescently coupled waveguides placed close together in parallel. Some research has described how to make a direction coupler active[35, 36]. Our idea is to use a silica zipper cavity to change the gap dynamically between the two nanowires that allow us to obtain different coupling strengths and hence switch the light direction. The operation of this micro sized mechanical system is driven by optical radiation force; hence we named it as a micro opto-mechanical system (MOMS). We would like to emphasize that there are numerous advantages to using a silica zipper cavity compared with cavities made of other materials. First, silica is a material with a wide bandgap that allows us to use wavelengths in the visible to telecom range. In this study we switch 1550-nm light with 770-nm light. The basic idea is that we design a cavity at 770 nm to allow efficient opto-mechanical coupling, but we design the device so that it is a directional waveguide for 1550-nm light. 1550-nm light corresponds to $0.22 a/\lambda$ in FIG. 1(c), which shows that the structure has negligible dispersion thus allowing us to transmit fast signals. Switching between two different wavelengths this far apart is not easy with other schemes. Secondly, silica is a low-loss material and our device has the potential for integration with sophisticated passive devices known as planar lightwave circuits. Thirdly, silica is a soft material (compared with silicon and silicon nitride), which may allow us to achieve radiation pressure assisted switching with a low control power.

To enable structural deformation at a small input, we redesigned the cavity length to 41.2 μm. We calculated the eigenfrequency of the in-plane differential mode and obtained $\Omega_m/2\pi = 3.55$ MHz and $m_{eff} = 14$ pg. To investigate the relationship between the gap distance and the extinction ratio of the device, we performed a 2D FDTD calculation. We used FDTD (rather than the beam propagation method) to confirm that there is only a small amount of reflection when light enters the photonic crystal. A schematic diagram of the computing model is shown in FIG. 6(a). A light source is placed to the upper left of the waveguide. The signal is telecom light whose wavelength is below the mode-gap of the cavity and that can propagate through the waveguide without exhibiting large dispersion. We observed the transmitted power at $e_1$, $e_2$, and $e_3$ in order to calculate the extinction ratio (ER). The ER is calculated with,

$$(ER) = 10\log(\frac{e_1}{e_2}) - 10\log(\frac{e_1}{e_3}) \tag{4}$$

where the values are 17.8 and -18.2 dB when the gaps are 194 and 93 nm, respectively. This means that we can obtain a sufficiently high switching contrast of ≥ 17.8 dB with about a 100-nm deformation of the structure.

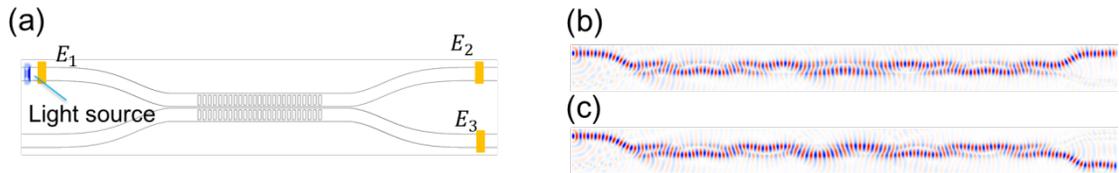

FIG. 6: (a) Schematic diagram of computing model. (b) Light propagation of the initial state, gap = 194 nm, ER = 17.8 dB. (c) Light propagation after deformation, $s$ = 93 nm, ER = 18.2 dB.

We use the bonded mode for driving the structure because it has a larger $g_{OM}$. Since it is an attractive force we start from 194 nm and reduce the gap to 93 nm. The boundary load (i.e. force) needed to move the structure for 100-nm can be calculated by FEM. The boundary area is set at 0.34 μm² (1.15$\lambda^2$), which is estimated from the mode volume of the cavity. After obtaining the force, we can calculate the required optical energy from FIG. 4(b). All these calculations are performed in COMSOL. Finally, we obtain the required input power $P_{in}$ from the energy $U$ of the light confined in the cavity ($U = QP_{in}/\omega$) since we know the $Q$ dependence versus the gap change. After all these calculations, we found that a 190-mW control light input is sufficient to move the structure and switch the light. This value is derived in a vacuum; however it should not be very different in air since it is not oscillating. In addition, our system has $Q_m = 2.0 \times 10^6$ in a vacuum, which determines the switching speed (i.e. the time required to reach a mechanically stable state). However, this value should be much smaller in air due to the additional damping.

Finally, we briefly consider the effect of regenerative amplification. During deformation, we have to keep the cavity static, so it is important to consider the oscillation threshold of the system. The mechanical damping ratio is given by the following equation[8, 37].

$$\Gamma_m' = \Gamma_m - \left(\frac{2U g_{OM}^2 \kappa}{\omega m_{eff}}\right) \Delta \left(\Delta^2 + \left(\frac{\kappa}{2}\right)^2\right)^2 \qquad (5)$$

Where $\Delta$ is laser detuning, $\kappa$ is the optical decay rate defined by $\kappa = \omega/Q$, and $\Gamma_m$ is the mechanical decay rate defined by $\Gamma_m = \Omega_m/Q_m$. For the calculation, we assume the mechanical quality factor determined by air damping to be $Q_m = 10$. The detuning dependence versus mechanical damping is shown in FIG. 5, when the gap is 94 nm, $g_{OM}/2\pi = 50$ GHz/nm, and $Q = 1.7 \times 10^4$. The structure oscillates when the normalized $\Gamma'/\Gamma$ has a minus value, so the regenerative threshold of this system is about 500 mW in a vacuum, well below the input power needed to move the structure.

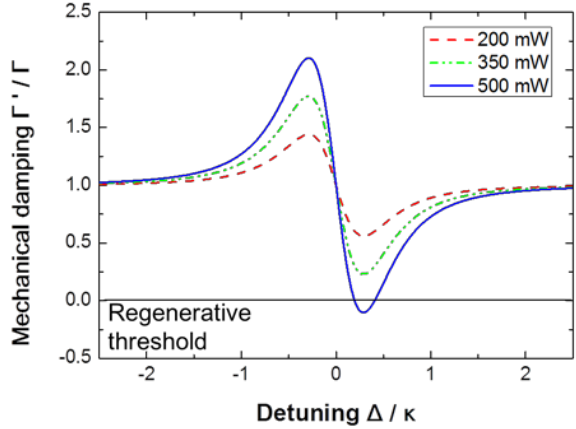

FIG. 7: Detuning dependence versus mechanical damping. Detuning is normalized by the optical decay rate and the mechanical damping rate is normalized by the intrinsic mechanical damping rate.

## V. CONCLUSION

We designed and analyzed a silica zipper cavity. The obtained optical quality factor was $Q = 6.0 \times 10^4$ and the effective mode volume $V_{\text{mode}}$ was $0.66\lambda^3$ when the slot gap was 34 nm. This value is five times higher than that of single silica nanobeam cavities. Then we show numerically that a strong optomechanical coupling rate as high as $g_{\text{OM}}/2\pi = 100$ GHz/nm is possible with this structure. Finally, we show that all-optical switching is possible by employing optical radiation pressure, and named such device as MOMS switch. We used two lights with very different wavelengths; i.e. 1550-nm telecom light is switched by using 770-nm visible light at a contrast of $\geq 17.8$ dB by using 190-mW input power. Although zipper cavities have been used for fundamental studies, we showed that a silica zipper cavity is also a promising platform for applications such as optical switches.